\documentclass{aa}  
\usepackage{graphicx,dblfloatfix,txfonts,xcolor,amsmath}
\begin{document} 

   \title{Alkali phenoxides in comets}
   \subtitle{}

   \author{M. Fulle\inst{1}
          \and
          P. Molaro\inst{1,2}
          \and
          A. Rotundi\inst{3}
          \and
          L. Tonietti\inst{3}
          \and
          A. Aletti\inst{4}
          \and
          L. Buzzi\inst{4}
          \and
          P. Valisa\inst{4}
          }

   \institute{INAF - Osservatorio Astronomico,
    Via Tiepolo 11, I-34143 Trieste, Italy\\
              \email{marco.fulle@inaf.it}
              \email{paolo.molaro@inaf.it}
        \and
        IFPU, Institute for Fundamental Physics of the Universe, Via Beirut 2, I-34151 Trieste, Italy
        \and
        Department of Science and Technology, Parthenope University of Naples, CDN-IC4, I-80143 Napoli, Italy
         \and
         Societ\`a Astronomica Schiaparelli,
    Via Giovanni Borghi 7, I-21100 Varese, Italy
             \\
             }

   \date{Received 25 February 2025; accepted 27 May 2025}

% \abstract{}{}{}{}{} 
% 5 {} token are mandatory
 
  \abstract
   {Potassium was first detected in spectra of the sungrazer comet C/1965 S1 Ikeya-Seki at the heliocentric distance $r_h = 0.15$ au and, 48 years later, in comets C/2011 L4 PanSTARRS and C/2012 S1 ISON at $r_h = 0.46$ au. The alkali tail photoionization model provides a Na/K ratio close to the solar value in comets C/1965 S1 and C/2011 L4. No lithium was detected in any comet: the lower limit of the Na/Li ratio was almost one order of magnitude greater than the solar ratio.}
   {Here we searched for the emissions of the alkali NaI, KI, and LiI in Comets C/2020 F3 NEOWISE and C/2024 G3 ATLAS.}
   {High-resolution spectra of the comets were taken with the 0.84 m telescope at the Schiaparelli Observatory at $r_h = 0.36$ and $r_h = 0.15$ au, respectively, the observations closest to the Sun since C/1965 S1. To model the data, we assumed that alkali phenoxides are present in the aromatic fraction of organic dust at the nucleus surface where they react with carbon dioxide ejecting alkali atoms.}
   {NaI and KI were detected in emission lines of exceptional intensity in both comets, with no evidence of LiI emission. The  NaI/KI ratios were determined: $31\pm 5$ and $26\pm 8$ in comets C/2020 F3 and C/2024 G3, respectively, whereas solar Na/K $\approx 15$. This excess and its observed trend with the heliocentric distance are consistent with chemistry between CO$_2$ and alkali phenoxides at the nucleus surface. The Li upper limit for comet C/2020 F3 is very stringent at Na/Li $> 3.4 ~10^4$, a factor of 34 greater than the solar value. This Li depletion is consistent with the reaction rate of lithium phenoxides, which is a factor of $10^4$ slower than sodium phenoxides.}
   {The widespread chemistry of carbon dioxide with organic dust may provide a significant energy and mass sink of carbon dioxide in all comets also at $r_h > 1$ au, reconciling recent models of cometary activity with Rosetta CO$_2$ measurements. At $r_h < 0.5$ au potassium was observed in all comets, so that we predict the formation of a KI tail spatially resolved from the NaI tail.}
   

   \maketitle
%
%________________________________________________________________

\section{Introduction}

Comets probably formed at low temperatures by the accretion of centimeter-sized pebbles \citep{blum2017} composed of dust particles \citep{levasseur2018} embedding ices \citep{fulle2020}. The derived structure of cometary nuclei \citep{ciarniello2022} is consistent with the dust ejection from nuclei \citep{fulle2020, fulle2022mnras} and with most of the data collected during the Rosetta mission \citep{ciarniello2023, fornasier2023}.

In this paper we focus on the alkalis Na, K, and Li in comets. While sodium has been observed in many comets \citep{combi1997, rauer1998, watanabe2003, leblanc2008, mckay2014}, potassium has been measured in the spectra of C/1965 S1 \citep{prreeston1967ApJ...147..718P}, C/2011 L4 \citep{fulle2013ApJ...771L..21F} and C/2012 S1 \citep{mckay2014}. The Na/K atomic ratio measured in C/1965 S1 and C/2011 L4 was much higher than the solar Na/K $= 15.6$ \citep{lodders2003}.

Lithium has never been detected in the spectra of comets. It is produced in the standard Big Bang nucleosynthesis but the value predicted once the baryon density is taken from the measure of primordial deuterium or from the cosmic background radiation is a factor of 3-4 higher than that measured in the old stars of the Galaxy. The sites for the lithium Galactic production remain elusive, though recently novae have been proposed as the most likely source (cf. \citet{molaro2023A&A...679A..72M} and references therein). A measure of lithium in comets could provide fresh clues to clarify one or both of these problems. The only ground-based LiI detection occurred during the impact of comet D/1993 F2 Shoemaker-Levy 9 on Jupiter \citep{roos-srote1995GeoRL..22.1621R}. The Na/Li ratio extracted from the spectra of the plume in Jupiter's atmosphere was consistent with a chondritic ratio Na/Li $= 1.04 ~10^3$ \citep{lodders2003}, although polluted by material from Jupiter's deep atmosphere \citep{costa1997}. Mass spectrum analyses of dust ejected by comet 81P/Wild 2 and returned by the Stardust mission allowed \citet{flynn2006} to determine Li, Na, and K abundances that are higher than the chondritic values.

\citet{fulle2013ApJ...771L..21F} considered the photoionization of the atoms ejected from the dust and flying from the sunward coma to the alkali tails observed in the slit of the spectrograph. This model implies a cross section of the alkali source (e.g., of optical thickness $10^{-6}$ for a loss rate of $10^3$ kg s$^{-1}$ of millimeter-sized dust) much smaller than that of the Moon, Mercury, and the nucleus of giant comets (e.g., C/1995 O1 Hale-Bopp), where sputtering extracts sodium efficiently \citep{combi1997}.

\begin{figure*}
   \centering
   \includegraphics[width=18.5truecm]{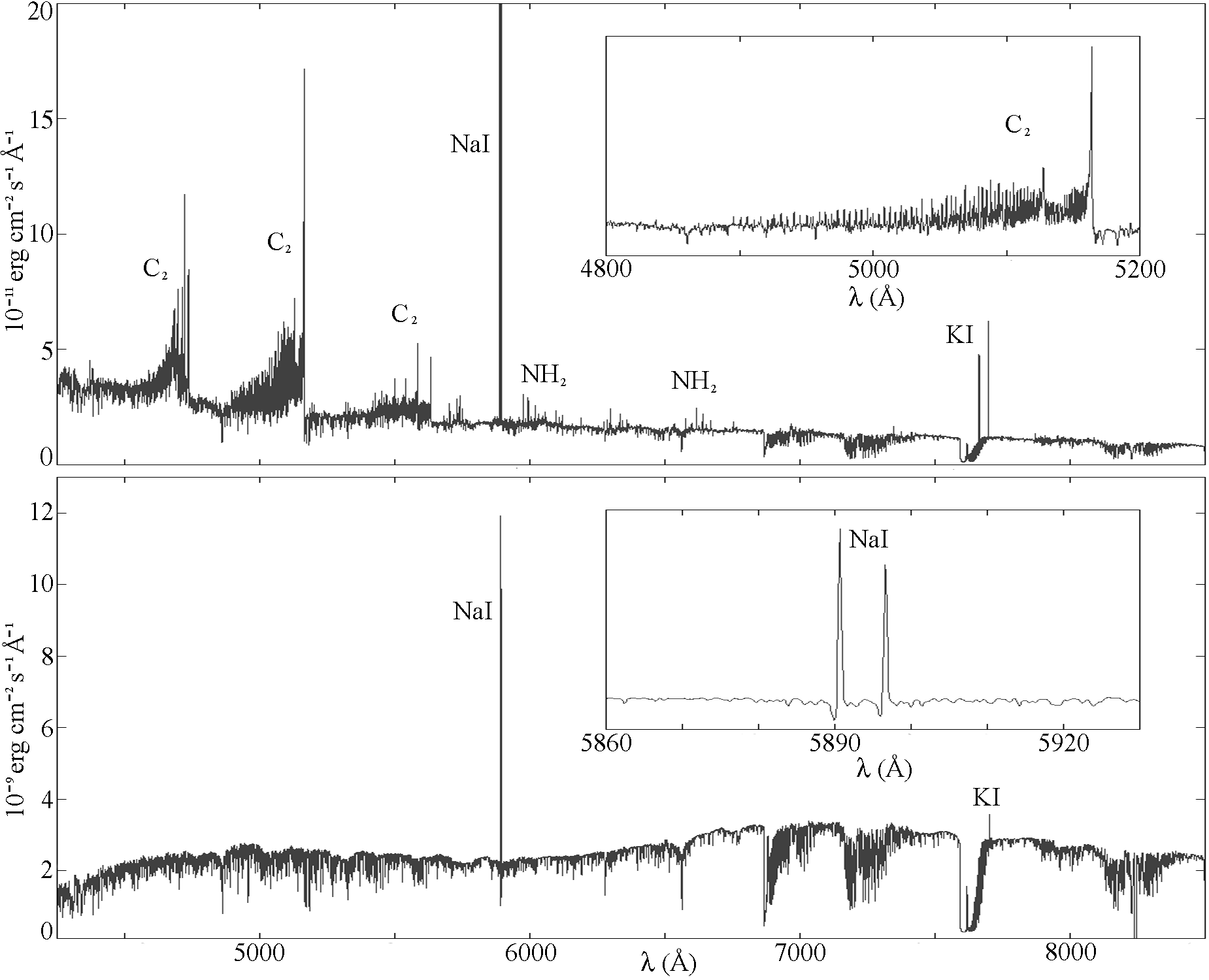}
   \caption{Spectra of comet C/2020 F3 NEOWISE (upper panel) taken on 2020 July 10.1 UT and of comet C/2024 G3 ATLAS (lower panel) taken on 2025 January 15.7 UT (observation log in Table \ref{tableobs}) shown over the complete observed wavelength range. Prominent features are the C$_2$ and NH$_2$ emission bands (upper panel), the NaI and KI emissions and the O$_2$ telluric absorption bands in the red part of the spectra.}
    \label{fig_spectra}
\end{figure*}

\begin{figure}
   \centering
   \includegraphics[width=8truecm]{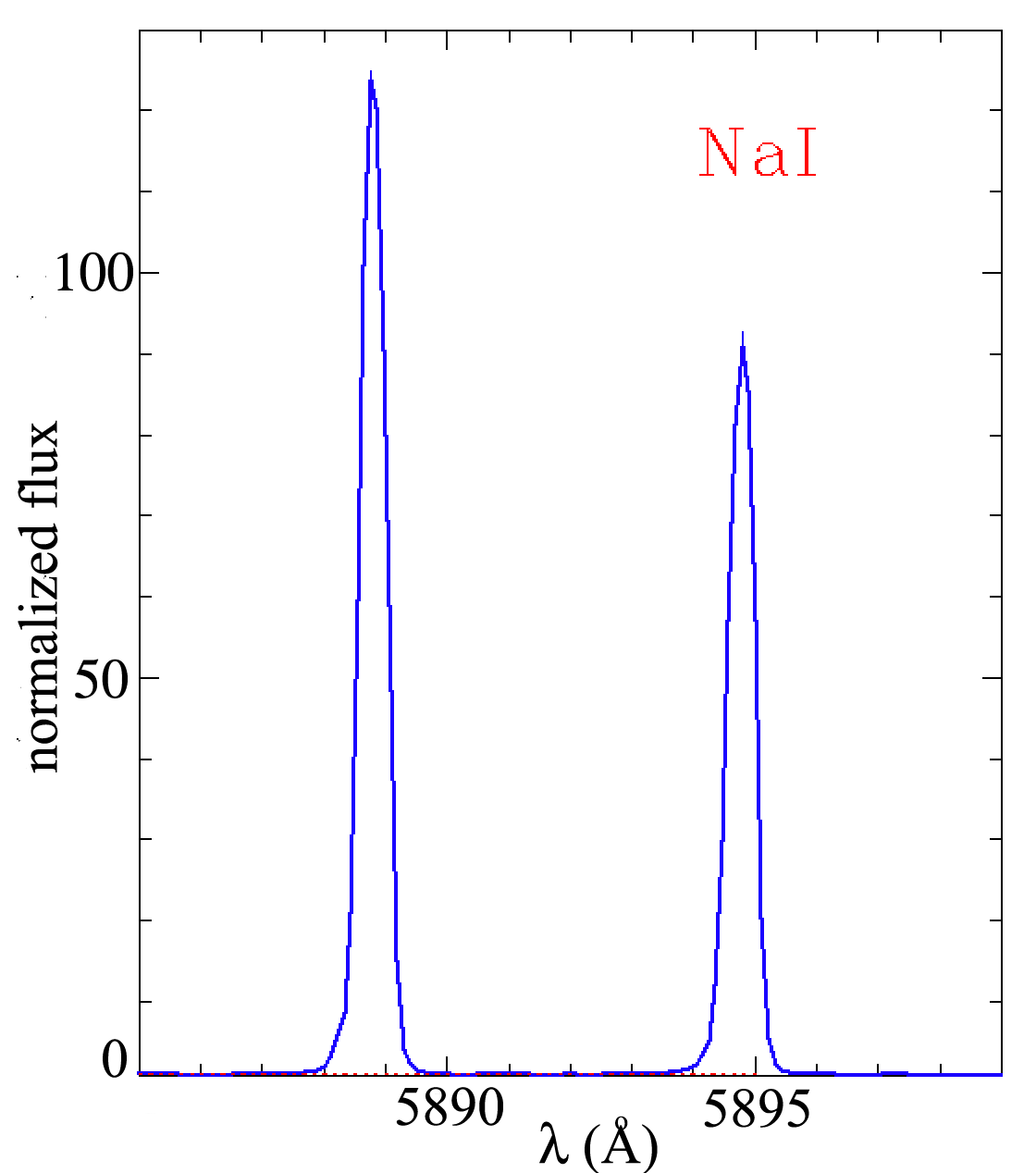}
   \caption{Spectrum of comet C/2020 F3 in the NaI 5890 \AA ~D line region. The flux has been normalized to $1.9 ~10^{-11}$ erg cm$^{-2}$ s$^{-1}$ \AA$^{-1}$.}
    \label{fig_neow10_na}
\end{figure}

\begin{figure}
   \centering
   \includegraphics[width=8truecm]{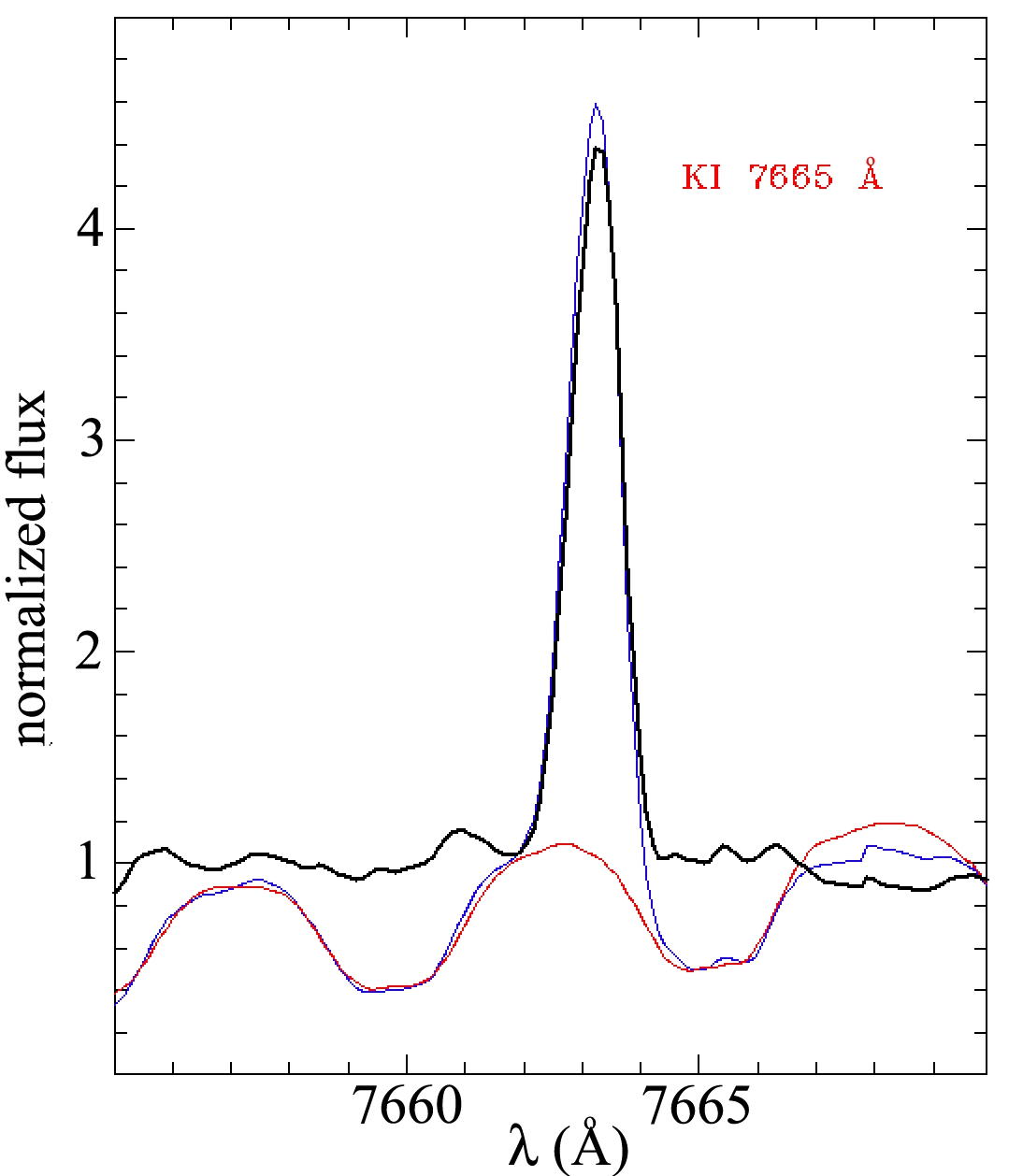}
   \caption{Spectrum of comet C/2020 F3 in the KI 7665 \AA ~line region. The observed spectrum (blue line) has been corrected (black line) for the sky contamination using the twilight spectrum (red line). The flux has been normalized to the twilight continuum of $1.04 ~10^{-11}$ erg cm$^{-2}$ s$^{-1}$ \AA$^{-1}$.}
    \label{fig_neowise_k7665}
\end{figure}

\begin{figure}
   \centering
   \includegraphics[width=9truecm]{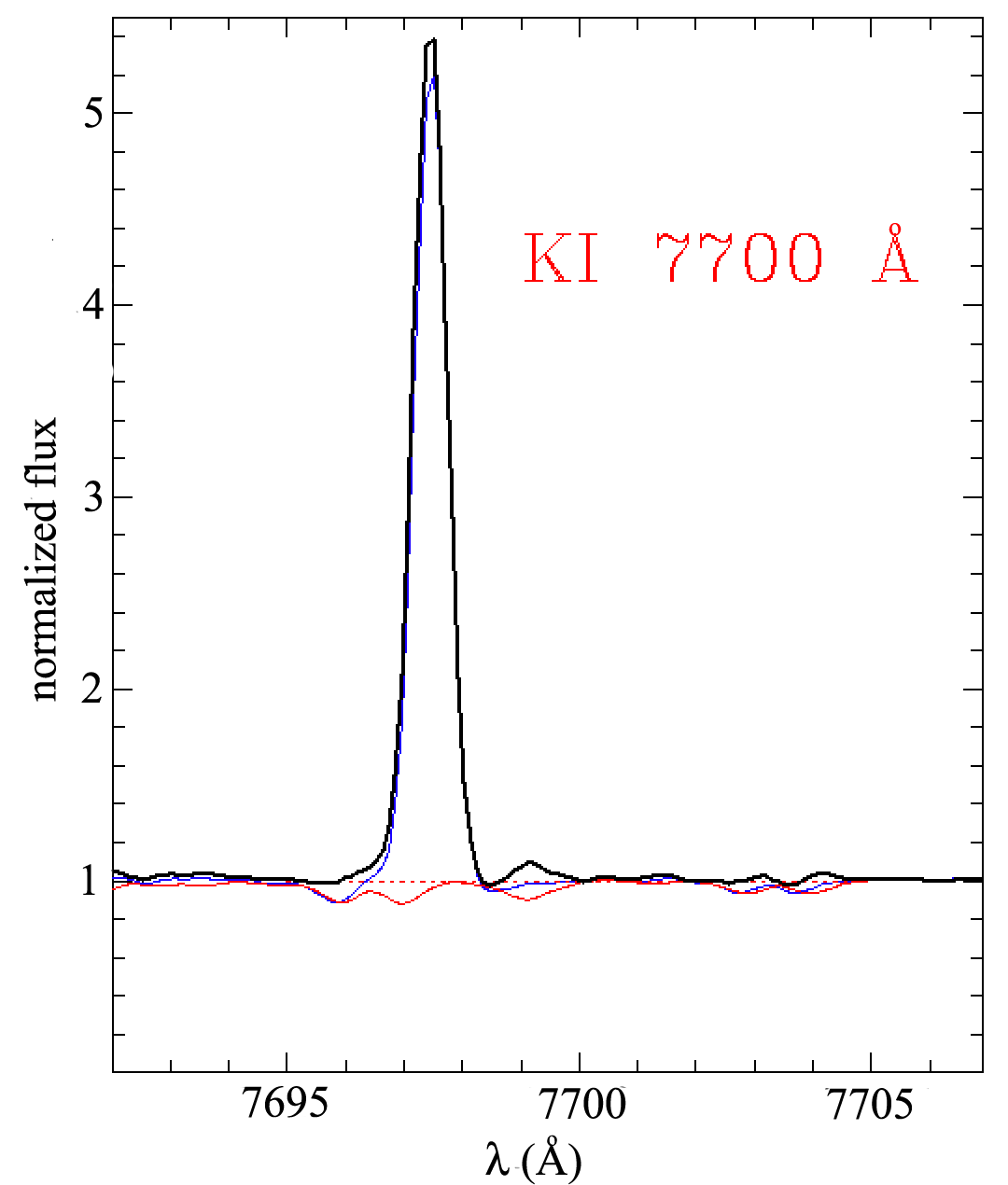}
   \caption{Spectrum of comet C/2020 F3 in the KI 7700 \AA ~line region. The observed spectrum (blue line) has been corrected (black line) for the sky contamination using the twilight spectrum (red line). The flux has been normalized to the twilight continuum of $1.2 ~10^{-11}$ erg cm$^{-2}$ s$^{-1}$ \AA$^{-1}$.}
   \label{fig_neowise_k7700}
\end{figure}

\begin{figure}
   \centering
   \includegraphics[width=9truecm]{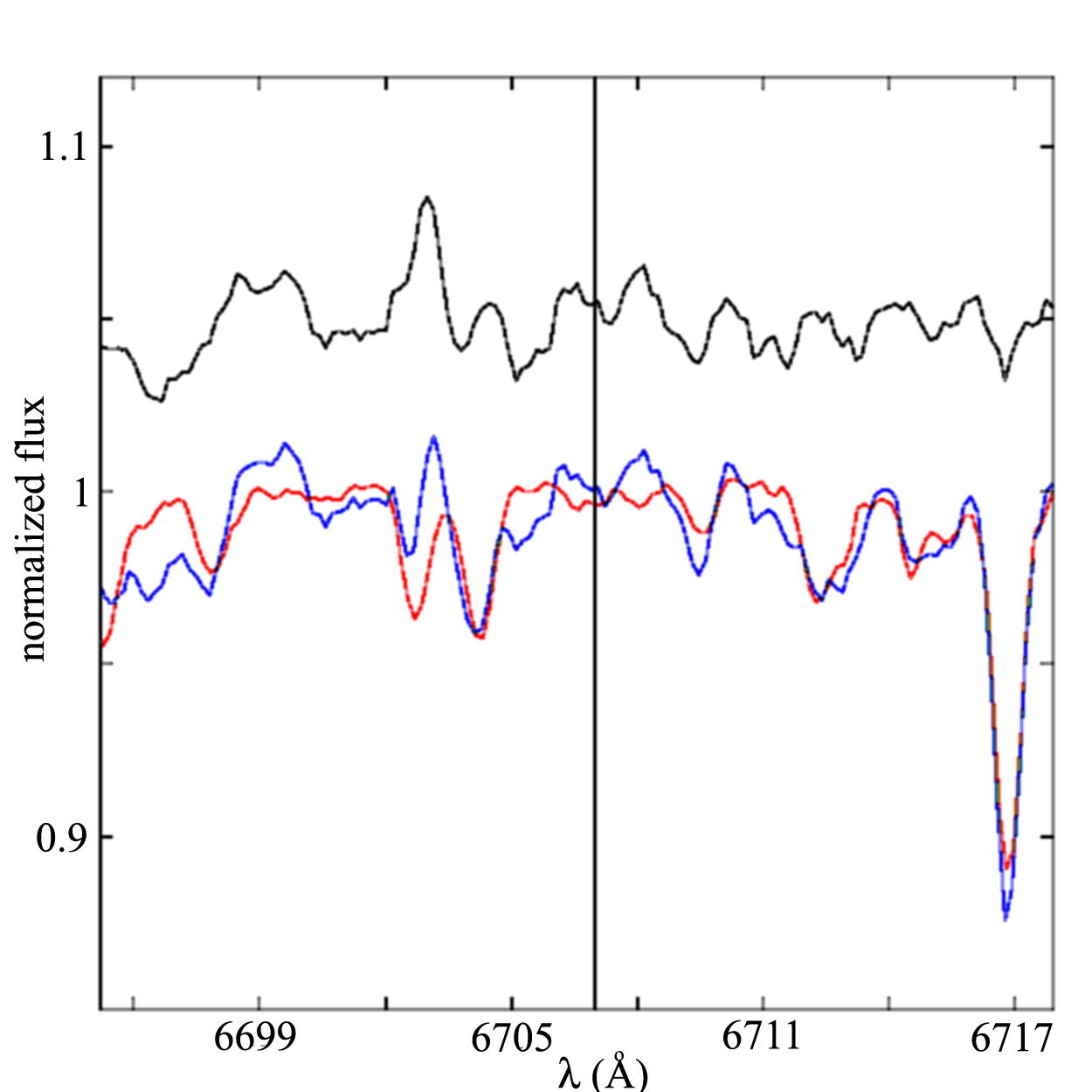}
   \caption{Spectrum of comet C/2020 F3 around the Li 6707 \AA ~line. The observed spectrum (blue line) has been corrected (black line, shifted up by 0.05) for the sky contamination using the twilight spectrum (red line). The flux has been normalized to $1.4 ~10^{-11}$ erg cm$^{-2}$ s$^{-1}$ \AA$^{-1}$.}
    \label{Fig_li_neowise}
\end{figure}

\begin{figure}
   \centering
   \includegraphics[width=8truecm]{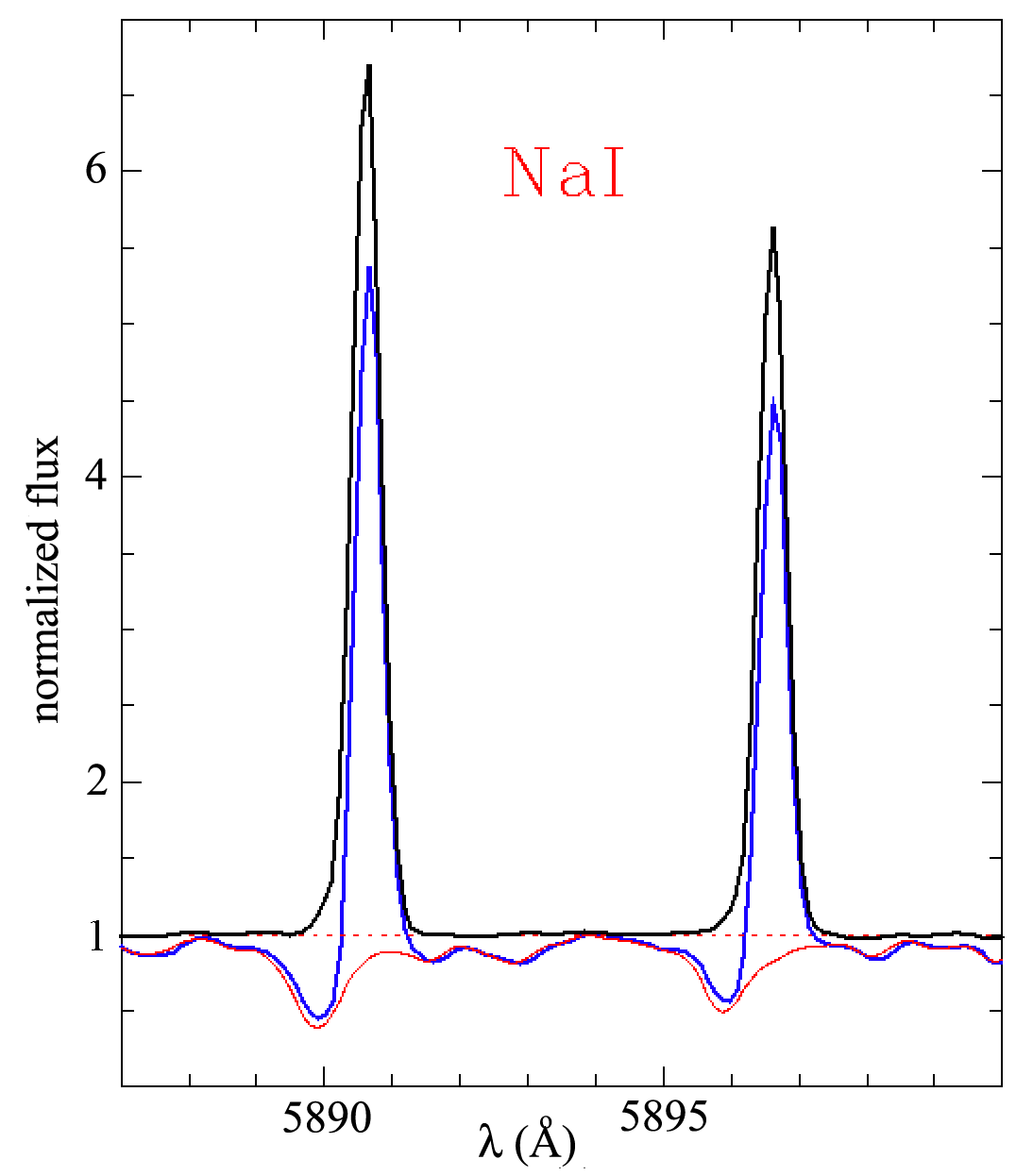}
   \caption{Spectrum of comet C/2024 G3 in the NaI 5890 \AA ~D line region. The observed spectrum (blue line) has been corrected (black line) for the sky contamination using the twilight spectrum (red line). The flux has been normalized to the twilight of $1.75 ~10^{-9}$ erg cm$^{-2}$ s$^{-1}$ \AA$^{-1}$.}
    \label{fig_atlas_na}
\end{figure}

\begin{figure}
   \centering
   \includegraphics[width=8truecm]{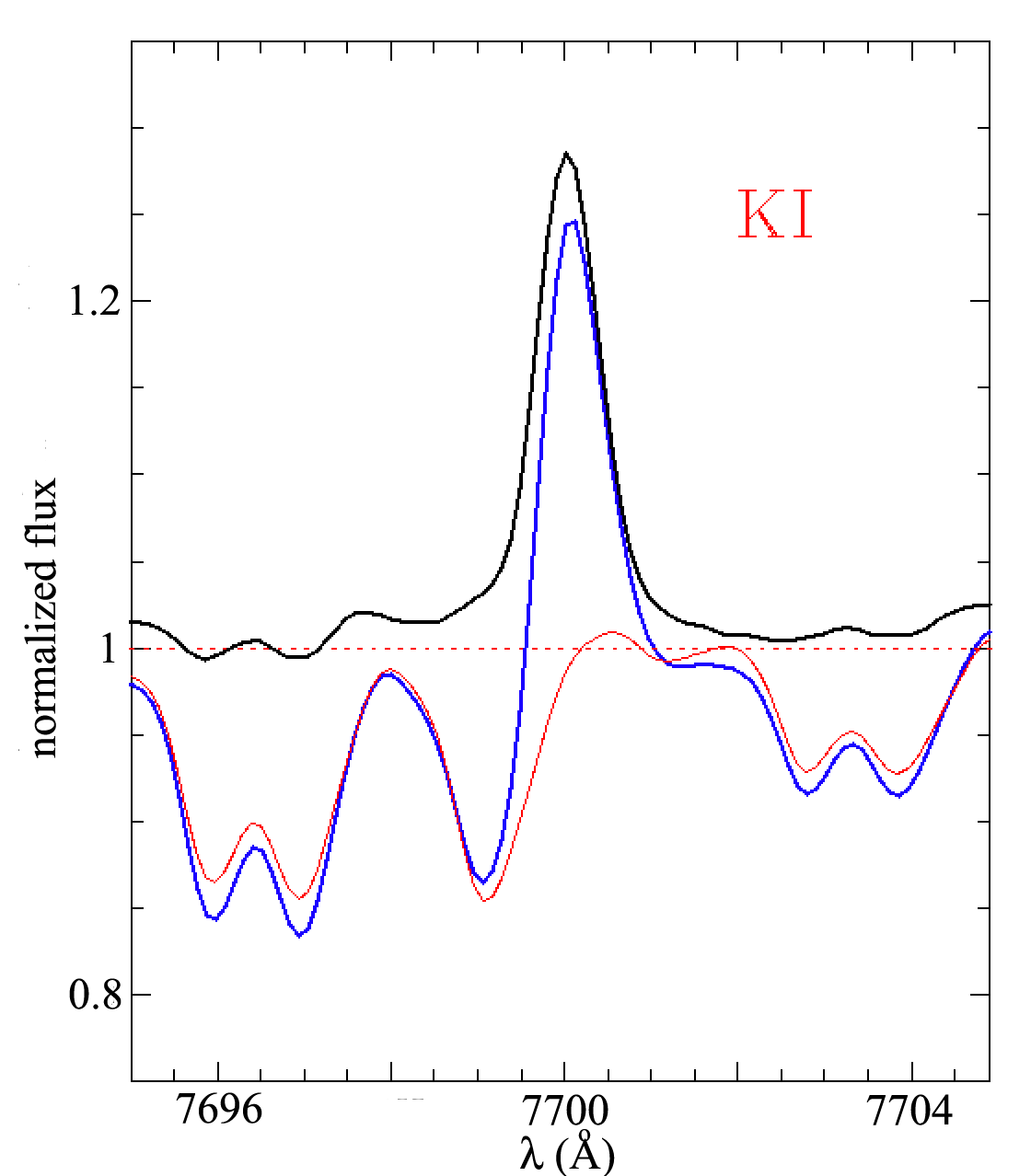}
   \caption{Spectrum of comet C/2024 G3 in the KI 7700 \AA ~line region. The observed spectrum (blue line) has been corrected (black line) for the sky contamination using the twilight spectrum (red line). The flux has been normalized to the twilight continuum of $2.84 ~10^{-9}$ erg cm$^{-2}$ s$^{-1}$ \AA$^{-1}$.}
    \label{fig_atlas_k7700}
\end{figure}

\begin{figure}[!h]
   \centering
   \includegraphics[width=8truecm]{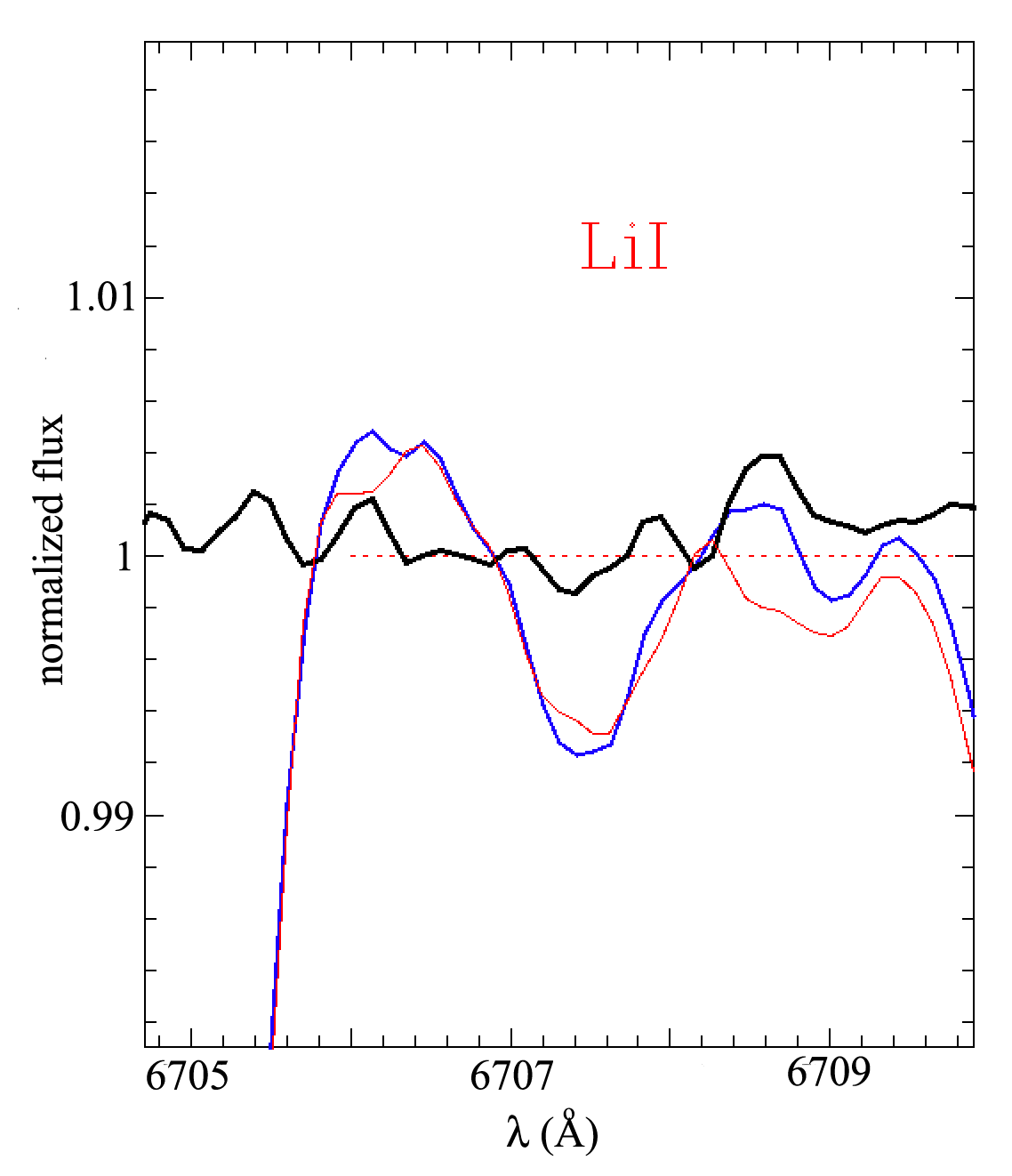}
   \caption{Spectrum of comet C/2024 G3 in the LiI 6707 \AA ~line region. The observed spectrum (blue line) has been corrected (black line) for the sky contamination using the twilight spectrum (red line). The flux has been normalized to the twilight of $3.1 ~10^{-9}$ erg cm$^{-2}$ s$^{-1}$ \AA$^{-1}$.}
    \label{fig_atlas_li}
\end{figure}

\begin{figure*}
   \centering
   \includegraphics[width=18.5truecm]{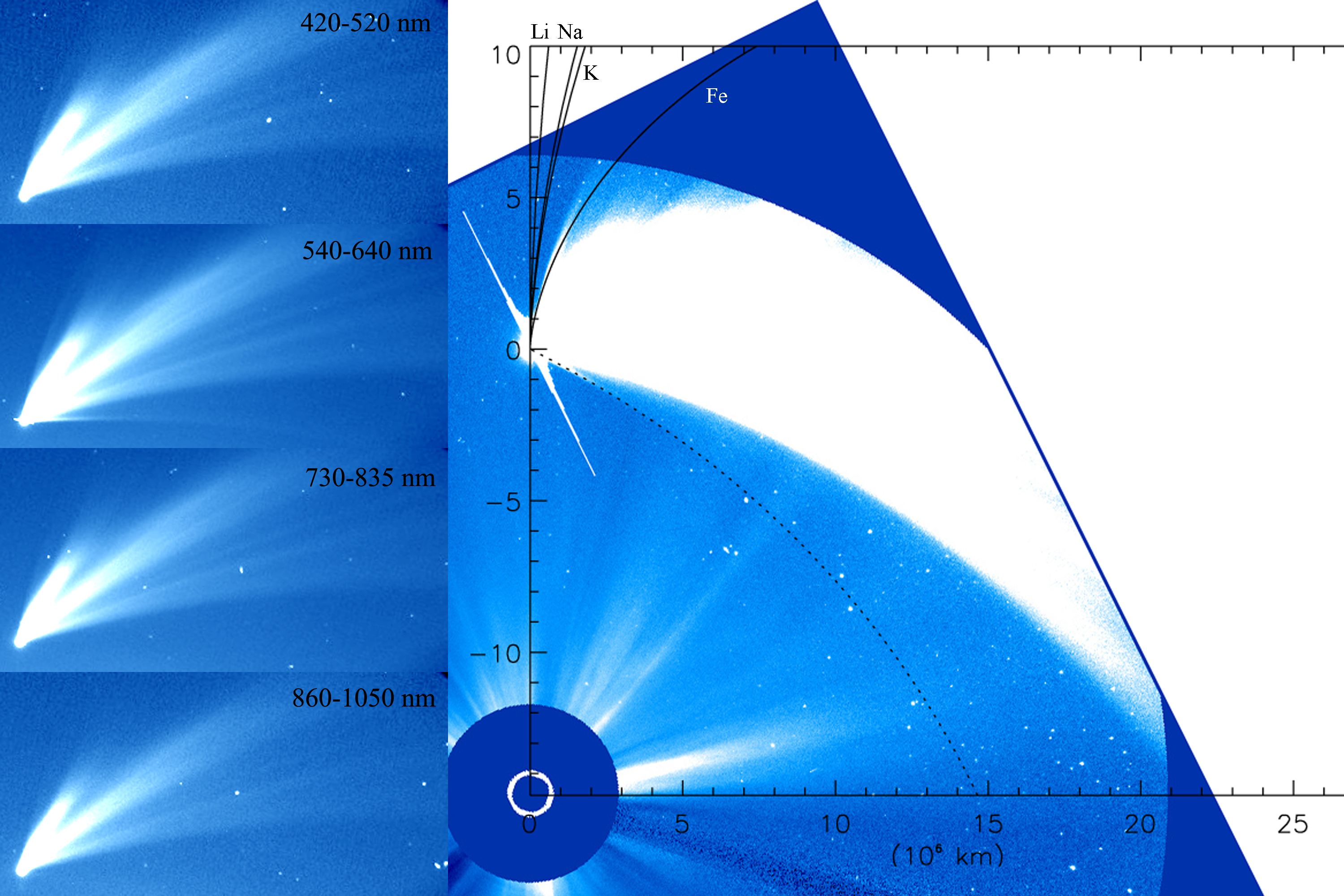}
   \caption{Left panels: C/2024 G3 ATLAS observed by the SOHO spacecraft on 2025 Jan 13.446-13.487 UT in four LASCO C3 passbands (Sun at bottom, images $1.2 ~10^7$ km wide). The horizontal trail is visible in the orange passband only (540-640 nm). Right panel: SOHO image taken on 2025 Jan 14.454 UT in the clear passband. The curved continuous lines are the predicted atomic neutral tails, from left to right: LiI tail (syndyne of $\beta = 440$); NaI tail (syndyne of $\beta = 71$); KI tail (syndyne of $\beta = 53$); FeI tail (syndyne of $\beta = 6$). Dotted line: Comet orbit fitting the trail.}
    \label{fig_tails}
\end{figure*}

\begin{table*}
\caption{Log of observations.}
\label{tableobs}
\centering
\begin{tabular}{ccccccccccccc}
\hline \hline
Comet &Time &$r_h$ &$\Delta$ &$dr_h/dt$ &$d\Delta/dt$ &Phase &Slit &Slit &Slit &Exp. &Air &Standard\\
&UT   &   &   &   &  &Angle &Width &Length &Area &Time &Mass &Star\\
& &au &au &km s$^{-1}$ &km s$^{-1}$ &deg &arcsec &arcsec &km$^2$ &sec. & & \\
\hline
C/2020 F3 &2020 Jul 10.1 &$0.36$ &$0.93$ &$+29$ &$-57$ &$94$ &$2$ &$17$ &$1.5 ~10^7$ &$600$ &$5$ &HR1558 \\
C/2024 G3 &2025 Jan 15.7 &$0.15$ &$0.96$ &$+67$ &$+40$ &$94$ &$3$ &$17$ &$2.5 ~10^7$ &$40 \times 5$ &$\approx 20$ & \\
\hline
\end{tabular}
\tablefoot{$r_h$ and $\Delta$ are the heliocentric and geocentric distances, respectively. The slit was north-south oriented.}
\end{table*}

\section{Observations}

Comets C/2020 F3 NEOWISE and C/2024 G3 ATLAS were observed with the Multi-Mode Spectrograph at the 0.84 m telescope of the Osservatorio Astronomico Schiaparelli in Campo dei Fiori, Varese, Italy \citep{munari2014}. The collected spectra are the first high-resolution ones of comets observed at $r_h < 0.4$ au since C/1965 S1 (Table \ref{tableobs}), and show strong NaI and KI lines (Figure \ref{fig_spectra}). The multi-order Echelle spectra cover the spectral range 424-864 nm at a resolving power $\lambda / \Delta\lambda \approx 10^4$.

Data reduction was performed in IRAF and included all the usual steps for bias, dark, flat, and wavelength calibration. Reference spectra with ArTh Hollow Cathode Lamp were recorded just before and after the science exposures.

\subsection{Comet C/2020 F3 NEOWISE}

The spectrum of C/2020 F3 NEOWISE (Figure \ref{fig_spectra}) was summed along the slit and shows a prominent C$_2$ Swan band, and NaI and KI emissions. Spectra of the standard star HR1558 were acquired before the spectra of C/2020 F3, and twilight spectra far from the comet just after. The NaI coma shows a Gaussian profile along the slit. No atmospheric NaI lines were detected in the sky spectrum. The spectrum of the solar light reflected by the cometary dust is much fainter than the twilight spectrum and was not detected. The exceptionally intense NaI $\lambda \lambda$ 5889.95 and $\lambda \lambda$ 5895.92 \AA ~lines (Figure \ref{fig_neow10_na}) were blueshifted by 59 and 58 km s$^{-1}$, respectively, consistent with the values in the ephemerides. 

The KI $\lambda \lambda$ 7664.90 and $\lambda \lambda$ 7698.96 \AA ~lines are shown in Figures \ref{fig_neowise_k7665} and \ref{fig_neowise_k7700}. The KI $\lambda \lambda$ 7664.90 \AA ~line shows the presence of the  strong telluric O$_2$ line at $\lambda \lambda$ 7664.73. Once corrected for the atmospheric absorption, the emissions were blueshifted with velocities of 65 and 59 km s$^{-1}$ in fair agreement with the sodium emission. The blueshift of the KI $\lambda \lambda$ 7664.90 \AA ~line is 6 km s$^{-1}$ higher than all the other lines (but only 20\% of the spectral resolution) and may be due to the pollution of the telluric O$_2$ line (Figure \ref{fig_neowise_k7665}), so that we measured the Na/K ratio on the other KI line (Figure \ref{fig_neowise_k7700}). No emission was detected at the LiI $\lambda \lambda$ 6707.78 \AA ~line (Figure \ref{Fig_li_neowise}).

\subsection{Comet C/2024 G3 ATLAS}

Comet C/2024 G3 was observed just after sunset during an extremely clear evening with dry atmosphere (humidity of 25\% with $3.7^\circ$ C). The sky brightness was overwhelming, and the comet was barely perceptible in the pointing and guiding cameras, whose exposure time was decreased to 64 microseconds to avoid saturation. Due to the manual guiding and the poor seeing, the NaI coma uniformly filled the slit. The twilight spectrum was acquired on the next day at exactly the same time and sky location. No spectrum of a standard star was possible at the huge airmass of the comet, so that flux calibration was done by means of archived spectra of standard stars taken with the same spectrograph at the highest airmasses. This procedure did not allow us to evaluate the error affecting the absolute flux, although it did not affect the precision of the intensity ratio between spectral lines.

The spectrum of C/2024 G3 ATLAS (Figure \ref{fig_spectra}) was summed along the slit and was characterized by prominent NaI emissions. The NaI $\lambda \lambda$ 5889.95 and 5895.92 \AA ~lines (Figure \ref{fig_atlas_na}) show a P Cygni profile due to a significant absorption at rest by the atmospheric daylight solar spectrum. When this was corrected by means of a twilight spectrum, the D2 and D1 emissions were redshifted by 34.1 and 34.7 km s$^{-1}$, respectively, lower than the values in the ephemerides by about 20\% of the spectral resolution. The KI $\lambda \lambda$ 7698.96 \AA ~line was also found in emission (Figure \ref{fig_atlas_k7700}). When the atmospheric absorption was subtracted, the emission was redshifted by 41.8 km s$^{-1}$, consistent with the values in the ephemerides. The KI $\lambda \lambda$ 7664.90 \AA ~line was completely absorbed by a strong telluric O$_2$ line at $\lambda \lambda$ 7664.73 and was too weak to be detected. No emission was detected at the LiI $\lambda \lambda$ 6707.78 \AA ~line (Figure \ref{fig_atlas_li}).

\subsection{Abundances}

The intensities or limits of the alkali emissions are reported for comet C/2024 G3 ATLAS (Table \ref{tableG3}) and for comet C/2020 F3 NEOWISE (Table \ref{tableF3}), shown according to increasing $r_h$ values.

The observed abundance of sodium related to an atom $x$, $Na/x_{obs}$ (Tables \ref{tableG3}, \ref{tableF3}, and \ref{tableL4}) depend on the g-factors $g_x$ at 1 au computed by \citet{fulle2013ApJ...771L..21F} and were obtained from the line intensity $I$ by means of the relationship
\begin{equation} 
(Na/x)_{obs} = (g_x/g_{Na}) (I_{Na}/I_x).
\end{equation}

\section{Potassium and lithium tails in comets}

We detected potassium in emission in comets C/2011 L4 PanSTARRS \citep{fulle2013ApJ...771L..21F}, C/2020 F3 NEOWISE, and C/2024 G3 ATLAS, so that we can predict that a potassium tail should be observable in future comets. The ratio of the solar radiation pressure to gravity forces, $\beta$, is higher for NaI than for KI, $\beta_{Na} = 71 \pm 8$, and $\beta_K = 53 \pm 2$ \citep{fulle2013ApJ...771L..21F}, so that the KI tail should lie close to the NaI tail though resolved at a slightly larger distance from the Sun-Comet prolonged radius vector. With respect to sodium, the acceleration is about 3/4 and the photoionization lifetime is 1/4. Therefore, the length of the potassium tail should be $\approx$ 1/5 of the sodium tail.

The Li tail, with $\beta_{Li} = 440 \pm 7$ \citep{fulle2013ApJ...771L..21F}, may never be observed because a lithium emission line has never been detected. The higher $\beta$ value for Li would place its tail close to the Sun-Comet prolonged radius vector. With respect to NaI, the acceleration is six times higher and the photoionization lifetime is 1/10 \citep{fulle2013ApJ...771L..21F}. Therefore, the length of the LiI tail should be approximately one-half of the length of the NaI tail. 

Precise sky-projected positions of the alkali and neutral iron \citep{fulle2007ApJ...661L..93F} tails in comet C/2024 G3, assuming that lithium is present, are plotted in Figure \ref{fig_tails}. Even though FeI lines are not detected in the C/2024 G3 spectrum, due to its low S/N, C/2024 G3 actually shows only the iron tail, probably due to the higher iron abundance, solar Fe/Na = 14.6 \citep{lodders2003}, and to the slower photoionization rate \citep{fulle2007ApJ...661L..93F}. FeI lines are observed from the UV to the IR spectrum, although the strongest lines are observed at $\lambda < 420$ nm. The first FeI tail was detected in C/2006 P1 at $630 < \lambda < 730$ nm \citep{fulle2007ApJ...661L..93F}. No dust particles may reach $\beta \approx 6$ \citep{fulle2007ApJ...661L..93F}. We encourage observers to detect these tails in the next coming bright comets with small perihelia by means of narrowband filters centered on the elemental transitions. The intensities of the KI and LiI tails relative to the NaI tail scale according to Eq. (1).

\section{Sodium trail}

\begin{figure}
   \centering
   \includegraphics[width=9truecm]{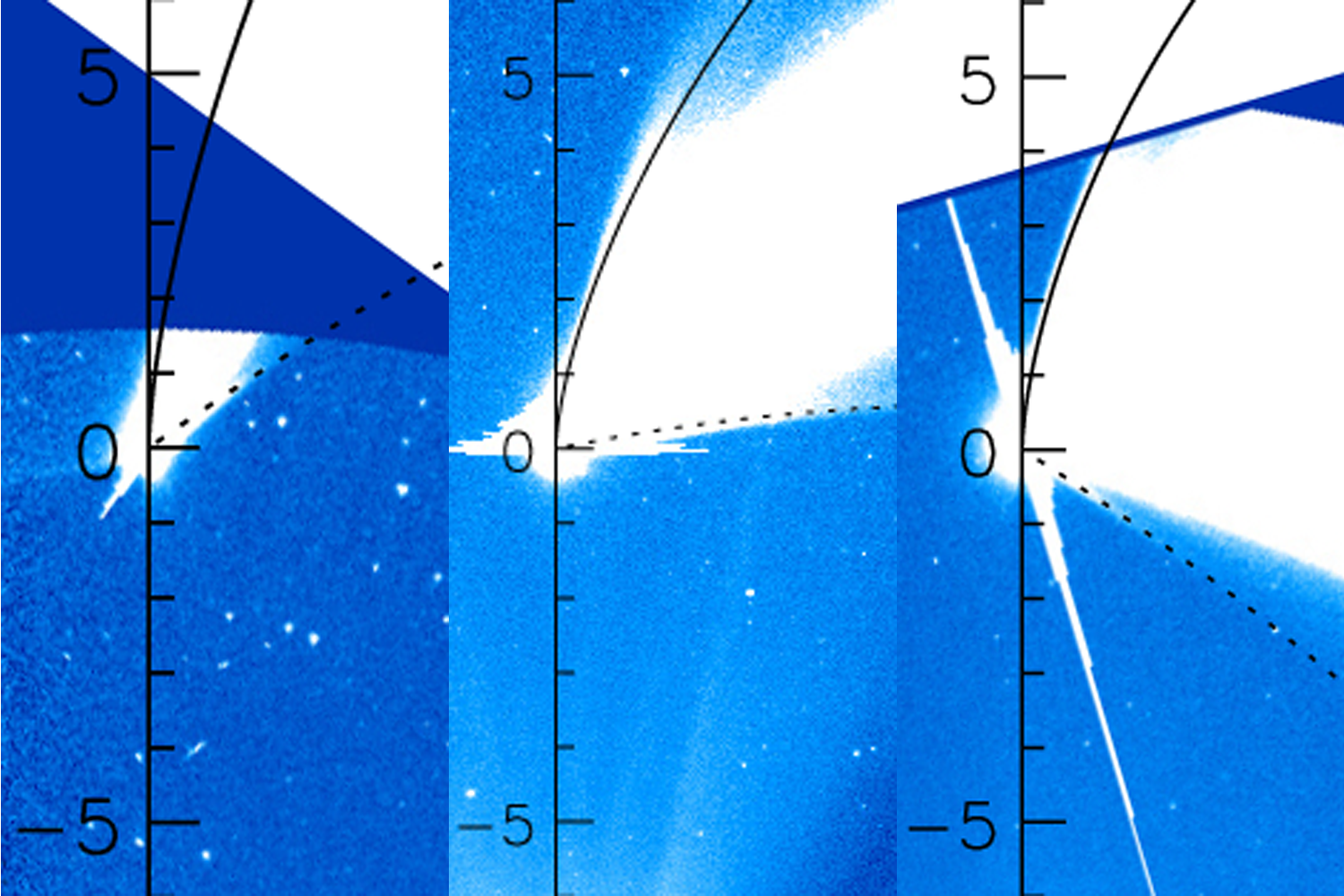}
   \caption{From left to right: C/2024 G3 ATLAS (ESA-NASA, SOHO) in the LASCO C3 clear passband observed on 2025 Jan 11.646, 13.204 and 14.746 UT, respectively. Vertical axis: Antisunward direction in $10^6$ km units. Curved line: FeI tail. Dotted line: Comet orbit fitting the trail.}
    \label{fig_trails}
\end{figure}

Contrary to the FeI and dust tails, which show a similar brightness in all the SOHO LASCO C3 passbands, the trail of C/2024 G3 (defined as a streamer of debris along the comet orbit, Figure \ref{fig_tails}) is the first ever observed in an Oort cloud comet and is visible in the orange passband only along an orbital arch covered by the comet in $\Delta T \approx 1$ day. The projection of the comet orbit on the sky observed by SOHO fits the trail in all SOHO images of C/2024 G3 (Figure \ref{fig_trails}). No trail is detected outside $540 < \lambda < 640$ nm, thus excluding solar light scattering by the trail debris and showing that the trail ejects sodium atoms. An ion tail (H$_2$O$^+$ bright band at $\lambda = 620$ nm) is excluded by the trail orientation.

The dynamics of the bodies building up the trail depend on the parameter $\beta$, which is linked to the diameter of the body $d$ by

\begin{equation} 
\beta = {{3 ~F_\odot} \over {8 \pi ~c ~G M_\odot ~\rho ~d}} = {C_{pr} \over {\rho ~d}},
\end{equation}

where $F_\odot$ and $M_\odot$ are the total solar radiation and mass, $c$ is the speed of light, $G$ is the gravitational constant, $C_{pr} = 1.19 ~10^{-3}$ kg m$^{-2}$ \citep{fulle2004come.book..565F}, and $\rho$ is the bulk density of the trail body [e.g., $\rho = 538$ kg m$^{-3}$ for the nucleus of comet 67P/Churyumov-Gerasimenko \citep{preusker2017}].

Two centuries ago, the osculating orbit of C/2024 G3 had the inbound parameters $q = 0.09445554$ au and $e = 0.99996987$. For $\beta \to 0$ and $e > 0$, the differences of the orbital parameters of the parent comet and of the body in the trail become

\begin{equation} 
\Delta e = \sqrt{1 + {{e^2 - 1 + {2 (1 + e) q \over r_s} \beta} \over {(1 - \beta)^2}}} - e = {{q(1 + e) \over r_s} - 1 + e^2 \over e} \beta + O(\beta^2),
\end{equation}
\begin{equation} 
\Delta q = {q(1 + e) \over {(1 + e + \Delta e)(1 - \beta)}} - q = \left(1 - {q \over r_s}\right) {q \over e} ~\beta + O(\beta^2),
\end{equation}

where $r_s$ is the heliocentric distance at which the trail body and the nucleus split [Eqs. (2b) and (2c) in \citet{fulle1989}]. The comet orbit fits the trail axis within one-half an image pixel $= 2 ~10^4$ km (Figure  \ref{fig_tails}), constraining the trail ejection either at the previous perihelion or at $r_s \approx q + \sqrt[3]{G M_\odot ~(10 ~\Delta T / \beta)^2} \ge 25$ au\footnote{If $e = 1$, $r_s = q (1 + f^2)$ and $\Delta T = k \beta \sqrt{q^3/G M_\odot} (f + {1 \over 3} f^3) + O(\beta^2)$: $f$ is the tangent of half true anomaly, $f \ll {1 \over 3} f^3$ for $5 \le r_s \le 500$ au, where, if $k = 0.3$, $r_s$ differs by $< 5$\% from precise computations done at $e < 1$.} because $e \to 1$, $\beta \le 1.4 ~10^{-3}$ (Eq. 4) and $\Delta e \le 10^{-5}$ (Eq. 3). The trail width implies ejection velocities $\le 0.1$ m s$^{-1}$, making the trail origin inconsistent with a cometary outburst \citep{jewitt2025}. The orbital period of C/2024 G3 is $T = [q/(1-e)]^{3/2}$ yr. The difference between the trail body orbital period and $T$ is

\begin{multline}
\Delta T = \sqrt {(q + \Delta q)^3 \over {(1 - e - \Delta e)^3(1 - \beta)}} - T = {{q^{3 \over 2} (1 - \beta)} \over {\left[1 - e - 2 {q \over r_s} \beta \right]^{3 \over 2}}} - T = \\ = \left[{3 \over {(1 -e)}}{q \over r_s} - 1\right] ~T ~\beta + O(\beta^2).
\end{multline}

If the trail was ejected at the previous perihelion, then $r_s = q$, i.e. $\Delta T \le 1$ kyr if $d \ge 30$ m: the trail would be too long and diluted to be detectable. If the trail was ejected at the previous aphelion $Q$, then $r_s = Q = q(1+e)/(1-e)$, i.e. $\Delta T /2 \le 1$ day as observed if $d \ge 30$ m. $Q$ lies in the Oort cloud, where some Galactic tides may have been strong enough to split the nucleus into the trail mininuclei. The perihelion solar tides definitely destroyed the main nucleus \citep{king2025}. At the previous aphelion, a Galactic tide of similar strength overcame the gravity bonds among the sub-kilometer-sized mininuclei, and decelerated C/2024 G3 from the Oort cloud orbital speed $\sqrt{G M_\odot / Q}$ to the inbound speed $\sqrt{(1-e) ~G M_\odot / Q}$ in a few years.

\begin{table}
%\tabletypesize{\scriptsize}
\caption{Atomic parameters in comet C/2024 G3 ATLAS, $r_h = 0.15$ au.}
\label{tableG3}
\centering
\begin{tabular}{cccccc}
\hline \hline
Atom &$\lambda$ (\AA) &$c {\Delta\lambda \over \lambda}$ &$I$ (3$\sigma$) &$(Na/x)_{obs}$ &$(Na/x)_{cor}$ \\
\hline
Li &6707.78 & &$\le$ 1.3 &$\ge 10^3$ &$\ge 250$ \\
Na &5889.95 &$+34$ &$532 \pm 5$ &  &  \\
Na &5895.92 &$+35$ &$438 \pm 5$ & 1 & 1 \\
K  &7664.90 & & &  &  \\
K  &7698.96 &$+42$ &$24 \pm 4$ &$26 \pm 8$  &$3 \pm 1$ \\
\hline
\end{tabular}
\tablefoot{Intensities $I$ are in $10^{-11}$ erg s$^{-1}$ cm$^{-2}$ \AA$^{-1}$ units. $c$ is the light velocity (km s$^{-1}$): the spectral resolution is $\approx 30$ km s$^{-1}$. The observed (obs) atomic ratios are provided by Eq. 1, the corrected (cor) ones by the alkali tail photoionization model \citep{fulle2013ApJ...771L..21F}.}
\end{table}

\begin{table}
\caption{Atomic parameters in C/2020 F3 NEOWISE, $r_h = 0.36$ au.}
\label{tableF3}
\centering
\begin{tabular}{cccccc}
\hline \hline
Atom & $\lambda$ (\AA) &$c {\Delta\lambda \over \lambda}$ & $I$ (3 $\sigma$) & $(Na/x)_{obs}$ & $(Na/x)_{cor}$ \\
\hline
Li &6707.78 & &$\le 0.009$ &$\ge 3.4 ~10^4$ & $\ge 1.2 ~10^4$ \\
Na &5889.95 &$-59$ &$120.8 \pm 0.5$ &  &  \\
Na &5895.92 &$-58$ &$88.1 \pm 0.5$ & 1 & 1 \\
K  &7664.90 &$-65$ &$3.78 \pm 0.05$ &  &  \\
K  &7698.96 &$-59$ &$3.80 \pm 0.05$ &$31 \pm 5$  & $8 \pm 1$ \\
\hline
\end{tabular}
\tablefoot{Intensities $I$ are in $10^{-11}$ erg s$^{-1}$ cm$^{-2}$ \AA$^{-1}$ units.}
\end{table}

\begin{table}
\caption{Atomic parameters in C/2011 L4 PanSTARRS, $r_h = 0.46$ au.}
\label{tableL4}
\centering
\begin{tabular}{ccccc}
\hline \hline
Atom & $\lambda$ (\AA) &$I$ (3 $\sigma$) & $(Na/x)_{obs}$ & $(Na/x)_{cor}$ \\
\hline
Li &6707.78 &$\le 0.027$ &$\ge 8 ~10^3$ & $\ge 4 ~10^3$ \\
Na &5889.95 &$85 \pm 1$ & 1 & 1 \\
Na &5895.92 &$49 \pm 1$ &  &  \\
K  &7664.90 &$2.1 \pm 0.2$ & $54 \pm 14$ & $18 \pm 5$ \\
K  &7698.96 &$1.3 \pm 0.1$ & & \\
\hline
\end{tabular}
\tablefoot{Intensities $I$ are in $10^{-11}$ erg s$^{-1}$ cm$^{-2}$ \AA$^{-1}$ \citep{fulle2013ApJ...771L..21F}.}
\end{table}

\section{Discussion}

Several processes are expected to extract alkali atoms from cometary nucleus and dust, namely thermal desorption, photon-stimulated desorption, and solar wind sputtering. From laboratory experiments on cosmic analogs it is expected that the atoms leaving the source maintain their original abundance \citep{leblanc2011Icar..211...10L}, and that the flux of ejected alkali is proportional to the source cross section. The trail of C/2024 G3 (Figures \ref{fig_tails} and \ref{fig_trails}) suggests that sodium ejection requires high temperatures and also a large volume of the source. In fact, the mininuclei in the trail eject more sodium than the dust tail, which has a much larger total cross section per image pixel.

 As for comets C/2011 L4 and C/1965 S1, the observed $Na/K_{obs}$ ratio for comets C/2020 F3 and C/2024 G3 (Tables \ref{tableG3} and \ref{tableF3}) is higher than the solar ratio, i.e. Na/K = 15.6 \citep{lodders2003}. The alkali tail photoionization model \citep{fulle2013ApJ...771L..21F} takes into account the alkali photoionization, by the solar UV radiation \citep{leblanc2011Icar..211...10L}, during their transfer into the alkali tails by solar radiation pressure, and provides $Na/K_{cor}$ values consistent with the solar ratio (Table \ref{tableL4}). However, the $Na/K_{obs}$ of C/2020 F3 and C/2024 G3 are corrected to $Na/K_{cor}$ values inconsistent with the solar ratio (Tables \ref{tableG3} and \ref{tableF3}).

Regarding comet C/2020 F3, high-resolution spectra allowed \citet{ye2020} to detect NaI and KI at $0.46 \le r_h \le 0.48$ au, and \citet{bischoff2021} and \citet{cambianica2021} to detect NaI at $0.6 \le r_h \le 0.7$ au, but not KI. Neither NaI nor KI was detected at $0.8 \le r_h \le 0.9$ au. Comet C/1996 B2 Hyakutake showed NaI emission but no KI emission at $0.5 \le r_h \le 0.7$ au \citep{hicks1997}. Comets 1P/Halley and 153P/Ikeya-Zhang showed NaI emission at $r_h < 0.8$ au only \citep{combi1997, watanabe2003}. Comet C/2012 S1 ISON showed NaI and KI emissions at $r_h = 0.46$ au \citep{mckay2014}. These facts, combined with our observations reported in Tables \ref{tableG3}, \ref{tableF3}, and \ref{tableL4}, allow us to conclude that sodium is ejected at $r_h < 0.8$ au, and potassium at $r_h < 0.5$ au, suggesting an ejection process that depends on the different temperatures of sodium and potassium rather than on sputtering or thermal desorption.

The only comet showing NaI emission at $r_h > 0.8$ au was C/1995 O1 Hale-Bopp, probably because its nucleus of radius $\approx 30$ km \citep{lamy2004} made efficient alkali sputtering from its surface. The bulk of its NaI emission line showed a sunward extension of $2 ~10^4$ km \citep{rauer1998}, consistent with the radius of a gas collisional coma \citep{zakharov2018}. Alkali sputtering at $r_h \approx 3$ au was detected by ROSINA when Rosetta was less than 10 km from the nucleus surface of 67P/Churyumov-Gerasimenko \citep{wurz2015}. A sodium source from dust-dust collisions requires dust nanograins \citep{ip1998}, excluded by the Rosetta mission \citep{levasseur2018}. \citet{leblanc2008} excluded thermal desorption as a significant sodium source by comparing the Na loss rates from C/1995 O1 and C/2006 P1. The sunward extension of C/2006 P1 NaI line of $\approx 10^4$ km \citep{leblanc2008} is consistent with a gas collisional coma if the nucleus radius is $\ge 10$ km, and the observed NaI emission along the trailing orbit may suggest a sodium trail much shorter than the C/2024 G3 trail.

The mass spectrum analysis of the craters in the aluminum foil exposed to the dust flux during the Stardust mission allowed \citet{flynn2006} to determine the elemental composition of silicates and organics ejected by comet 81P/Wild 2; they found lithium, sodium, and potassium in higher abundances with respect to the chondritic values. The fact that lithium was never detected in the spectra of comets (e.g., in comet C/2020 F3 characterized by a Na/Li 34 times higher than the solar value) suggests the need of a new source of the observed alkali atoms consistent with different activation temperatures and relative abundances.

Here we assume that a fraction of sodium and potassium is contained in the aromatic component of cometary dust \citep{sandford2006}, namely in phenoxides \citep{kojcinovic2024}, where sodium and potassium oxides are bonded to a benzene ring. This assumption is consistent with the Na/Fe abundances that are higher than the chondritic value, as measured in dust of comets 1P/Halley by Giotto, in 81P/Wild 2 by Stardust, and in 67P/Churyumov-Gerasimenko by Rosetta \citep{flynn2006, levasseur2018}. Giotto and Rosetta, unlike Stardust, found a chondritic abundance of potassium, suggesting that the potassium fraction in phenoxides may be lower than that of sodium by a factor of two to three, directly matching $Na/K_{obs}$.

\begin{table}
\caption{Nucleus parameters according to the WEB model.}
\label{tableWEB}
\centering
\begin{tabular}{cccc}
\hline \hline
$r_h$ (au) & $T_s$ (K) & $\nabla T$ (K cm$^{-1}$) & $E$ (cm day$^{-1}$) \\
\hline
0.2 &585 &470 &32 \\
0.3 &480 &300 &28  \\
0.5 &400 &180 &23 \\
0.8 &340 &100 &18 \\
\hline
\end{tabular}
\end{table}

Laboratory experiments have shown that carbon dioxide extracts sodium and potassium atoms from phenoxides forming salicylic acid \citep{kojcinovic2024}. In particular, at CO$_2$ pressures higher than 1 MPa, phenoxides absorb carbon dioxide ejecting sodium atoms in the temperature range $350 < T < 450$ K, while potassium atoms are ejected for $470 < T < 490$ K. The observed increase in $Na/K_{obs}$ as $r_h$ increases (Tables \ref{tableG3}, \ref{tableF3}, and \ref{tableL4}) is consistent with the higher reaction temperature of potassium phenoxides with respect to that of sodium phenoxides. The mininuclei observed in the trail of C/2024 G3 are big enough to maintain CO$_2$ ice below their surface. The NaI line profile of C/2020 F3 along the slit of the spectrograph has a FWHM $\approx 4 ~10^3$ km, fitting the diameter of a gas collisional coma if the nucleus radius is $\ge 2$ km \citep{zakharov2018}. The NaI line intensities (Table \ref{tableF3}) provide a C/2020 F3 loss rate of $\approx 3 ~10^{25}$ Na s$^{-1}$, to be compared to the water loss rate of $4.8 ~10^{29}$ mol s$^{-1}$ at $r_h = 0.34$ au \citep{faggi2021}. The C/2020 F3 Na loss rate is a factor $\ge 30$ smaller than that from C/2006 P1 \citep{leblanc2008}, fitting the squared ratio of the respective nucleus radii.

The water enriched blocks (WEB) model of cometary nuclei \citep{fulle2020, ciarniello2022} provides the surface temperatures $T_s$ and thermal gradients $\nabla T$ inside the centimeter-sized pebbles at the nucleus surface (Table \ref{tableWEB}), which are perfectly consistent with the temperatures required by the reaction between CO$_2$ and alkali phenoxides: potassium phenoxides eject potassium atoms at $470 < T < 490$ K \citep{kojcinovic2024}, i.e., at $r_h < 0.5$ au (Table \ref{tableWEB}), and sodium phenoxides eject sodium atoms at $350 < T < 450$ K \citep{kojcinovic2024}, i.e., at $r_h < 0.8$ au (Table \ref{tableWEB}), as actually observed. The full temperature range $350 < T < 450$ K is reached at $r_h < 0.5$ au (Table \ref{tableWEB}), directly explaining the steep $r_h$ dependence of the sodium loss rate observed at $0.5 < r_h < 0.8$ au \citep{watanabe2003}. According to the WEB model  \citep{fulle2022mnras}, most of the carbon dioxide is present in the water-poor matrix embedding the WEBs  where the draining of ices $D$ is faster than the nucleus erosion $E$ \citep{fulle2020}, suggesting that the proposed chemical reaction may last from 45 to 80 minutes (Table \ref{tableWEB}). At the corresponding reaction temperatures, within these times about half of the phenoxides are converted into salicylic acid  \citep{kojcinovic2024}. During the C/2024 G3 perihelion week, the erosion $E$ reaches 2 m, exposing the nucleus pristine interior as well as its disintegration does. 

At a given temperature, the water vapor pressure is thousands of times lower than that of carbon dioxide, thus excluding chemical reactions between water vapor and alkali phenoxides. On the other hand, alkali phenoxides easily react with terrestrial liquid water \citep{kojcinovic2024}, making their presence in CI chondrites improbable, although phenol-derivates in meteorites can be generated through Fischer-Tropsch type reactions in the solar nebula \citep{hayatsu1980}. No specific laboratory analysys in search of phenoxides was performed on the Stardust samples \citep{sandford2006}. Theoretical analysis demonstrates the potential presence of phenol-like molecules in the interstellar medium \citep{ashworth2021, ghosh2022}.

At the pebble temperatures listed in Table \ref{tableWEB}, the CO$_2$ pressure is always higher than 1 MPa \citep{fulle2022mnras}, without implying a significant dust ejection because $D > E$. This is consistent with the apparent lack of dust tails associated with the mininuclei in the sodium trail of C/2024 G3. The proposed chemical reaction is a significant CO$_2$ energy and mass sink, which may avoid CO$_2$ pressure overcoming values of a few megapascals. Alkali atoms should be extracted from homogeneous porous aggregates of monomer grains 0.1 $\mu$m in size \citep{mannel2019}, i.e., dust particles that may survive at the megapascal pressures of CO$_2$ inside them \citep{blum2017}. Fireball observations suggest internal tensile strengths from 0.01 to 10 MPa in meteoroids associated with comet 2P/Encke \citep{wetherill1982}.

Lithium was not detected in the spectra of four comets where it looked for. The most stringent upper limit was derived for comet C/2020 F3 where Li/Na is found to be a factor of 34 below the solar value. Assuming a solar Na abundance, this ratio implies a lithium abundance of $A(Li) < 1.75$, where $A(Li) = \log (Li/H) + 12$. This is at variance with the Stardust samples where it is more abundant than in CI chondrites, namely $A(Li) > 3.28$ \citep{flynn2006}. The derived upper limit is one order of magnitude lower than the $A(Li) \approx 2.7$ predicted by theoretical primordial nucleosynthesis and also lower than the $A(Li) \approx 2.2$ observed in the dwarf stars of the Galactic halo \citep{navas2024}. Lithium in comets may be present either in silicates only, or in phenoxides reacting with carbon dioxide according to different reaction pathways \citep{markovic2007} characterized by an Arrhenius factor $10^4$ times lower than the reactions of sodium phenoxide \citep{staude1971}.

\section{Conclusions}

   \begin{enumerate}
   
     \item{We analyzed the first high-resolution spectra of comets observed at heliocentric distances $r_h < 0.4$ au since C/1965 S1, showing the strongest NaI and KI lines ever recorded.}

      \item{The excess of the Na/K ratio with respect to the solar value and its observed trend with the heliocentric distance are consistent with alkali phenoxides in the aromatic fraction of cometary dust \citep{sandford2006} reacting with CO$_2$ to eject sodium and potassium atoms \citep{kojcinovic2024}.}

       \item{Lithium has never been detected in the spectra of comets. The Li upper limit for comet  C/2020 F3 is very stringent at a factor of 34 below the solar value, at variance with the Stardust samples where it is more abundant than in CI chondrites \citep{flynn2006}. The Na/K ratio is always greater than the solar ratio by a factor of two to three. KI is detected at $r_h < 0.5$ au, whereas NaI at $r_h < 0.8$ au. These three facts exclude thermal, photon-stimulated, and solar-wind desorptions, which should maintain cosmic abundances.} 
       
      \item{Lithium may be present either in silicates only, or in phenoxides reacting with carbon dioxide at a rate $10^4$ times slower than sodium phenoxides \citep{staude1971}.}
      
      \item{The widespread chemistry of carbon dioxide with organic dust \citep{Liu2015} may provide a significant energy and mass sink of carbon dioxide in all comets, also at heliocentric distances $> 1$ au, reconciling models of cometary activity \citep{Attree2024b, Attree2024a} with Rosetta CO$_2$ measurements.}
      
      \item{For comets observed at heliocentric distances $< 0.5$ au, we predict a potassium neutral tail spatially resolved from the sodium tail, and encourage its observation.}
      
      \item{C/2024 G3 is the first Oort cloud comet to show a trail, detected thanks to its sodium emission and possibly composed of sub-kilometer-sized mininuclei fragmented in the Oort cloud by the Galactic tides that decelerated C/2024 G3.}
      
\end{enumerate}

\begin{acknowledgements}
      We thank an anonymous referee for having significantly improved a previous version of the manuscript. We thank J. Agarwal, R. Ligustri, J. Markkanen and F. Moreno for useful discussions about the C/2024 G3 trail. We are grateful to A. Milani for quickly modifying the pointing camera acquisition software to reduce the exposure time thus avoiding saturation. Part of this work was supported by the ASI-INAF agreement 2023-14-HH.0 project number.
\end{acknowledgements}

\bibliography{aa54255-25.bib}

\end{document}